\documentclass[10pt,conference]{IEEEtran}
\IEEEoverridecommandlockouts

\usepackage{bbding}

\def\BibTeX{{\rm B\kern-.05em{\sc i\kern-.025em b}\kern-.08em
    T\kern-.1667em\lower.7ex\hbox{E}\kern-.125emX}}

\usepackage{colortbl}
\usepackage{latexsym}
\usepackage{color}
\usepackage{listings}
\usepackage{xcolor}
\usepackage{mciteplus}
\usepackage{graphicx}
\usepackage{amsmath}
\usepackage{amsthm, bm}
\usepackage{tabulary}
\usepackage{multirow,makecell}
\usepackage{booktabs}
\usepackage{soul}
\usepackage{hyperref}
\usepackage{bbold}
\usepackage{subfigure}   
\usepackage{bbding}
\usepackage{enumitem}
\usepackage{caption}
\usepackage{textcomp}  
\usepackage{scalerel}  
\usepackage{textcomp}  
\usepackage{flushend}

\usepackage[a-1b]{pdfx}

\usepackage[ruled,linesnumbered]{algorithm2e}

\pdfmapline{pdftex_dl14.map}

\newtheorem{definition}{Definition}

\renewcommand{\baselinestretch}{1.0}

\usepackage{microtype}

\usepackage{pifont}
\usepackage[noadjust]{cite}

\newcommand{\Rom}[1]{(\uppercase\expandafter{\romannumeral #1\relax})}

\definecolor{green}{HTML}{C1F8CF}
\definecolor{best}{HTML}{C1F8CF}
\definecolor{second}{HTML}{9ED8DB}
\definecolor{third}{HTML}{6D9F71}

\newcommand{\bestcell}{{\cellcolor{best}}}

\newcommand{\response}[1]{\textcolor{black}{#1}}
\newcommand{\rp}[1]{\textcolor{black}{#1}}

\newcommand{\sm}{Supplementary Material~\cite{snapshot}}

\newcommand{\F}{Fig.}
\newcommand{\E}{Eqn.}
\newcommand{\T}{Table}
\renewcommand{\S}{Sec.}
\newcommand{\A}{Alg.}

\newcommand{\ignore}[1]{}

\newcommand{\parh}[1]{\noindent\textbf{#1}}
\newcommand{\sparh}[1]{\noindent\underline{#1}}

\newcommand{\tool}{\textsc{PerfCE}\xspace}
\newcommand{\cm}{Chaos Mesh}

\newcommand{\CBrush}{\textcolor[RGB]{84,130,53}{\Checkmark}}
\newcommand{\XBrush}{\textcolor[RGB]{176,35,24}{\XSolidBrush}}

\begin{document}

\title{\tool: Performance Debugging on Databases with Chaos
Engineering-Enhanced Causality Analysis}

\author{
    \IEEEauthorblockN{
        Zhenlan Ji,
        Pingchuan Ma\IEEEauthorrefmark{1},
        and Shuai Wang
    }\thanks{\IEEEauthorrefmark{1} Corresponding authors}
    \IEEEauthorblockA{
        The Hong Kong University of Science and Technology,
        \{zjiae, pmaab, shuaiw\}@cse.ust.hk
    }
}

\maketitle

\begin{abstract}

Debugging performance anomalies in databases is challenging. Causal inference
techniques enable qualitative and quantitative root cause analysis of
performance downgrades. Nevertheless, causality analysis is challenging \rp{in
practice}, particularly due to limited \textit{observability}. Recently, chaos
engineering (CE) has been applied to test complex software systems. CE
frameworks mutate chaos variables to inject catastrophic events (e.g., network
slowdowns) to stress-test these software systems. The systems under chaos stress
are then tested (e.g., via differential testing) to check if they retain normal
functionality, such as returning correct SQL query outputs even under stress.

To date, CE is mainly employed to aid software testing. This paper identifies
the novel usage of CE in diagnosing performance anomalies in databases. Our
framework, \tool, has two phases --- offline and online. The offline phase
learns statistical models of a database using both passive observations and
proactive chaos experiments. The online phase diagnoses the root cause of
performance anomalies from both qualitative and quantitative aspects on-the-fly.
In evaluation, \tool\ outperformed previous works on synthetic datasets and is
highly accurate and moderately expensive when analyzing real-world (distributed)
databases like MySQL and TiDB.


\end{abstract}




%

\section{Introduction}
\label{sec:introduction}

Databases are critical infrastructures that support daily operations and
businesses. Service outages or performance defects can result in a negative user
experience, a decline in sales, and even brand damage. Google, for instance,
assesses page speed for ranking websites~\cite{wire19}. According to reports,
every 100ms of latency \rp{costs} Amazon 1\% in revenue~\cite{gigaspace}, and
every 0.5s of additional load delay for Google search results \rp{leads to} a
20\% loss in traffic~\cite{mayer}. Modern databases often entail complex
resource management, and dependencies between modules of a (distributed)
database may introduce subtle performance bottlenecks and degrade system
throughput. Diagnosing performance issues is cumbersome and error-prone,
especially as typical (distributed) databases on the cloud or containerization
environments may be exposed to hundreds of potentially influencing key
performance indicators (KPIs).

\begin{figure}[t]
\centering
\includegraphics[width=1.00\linewidth]{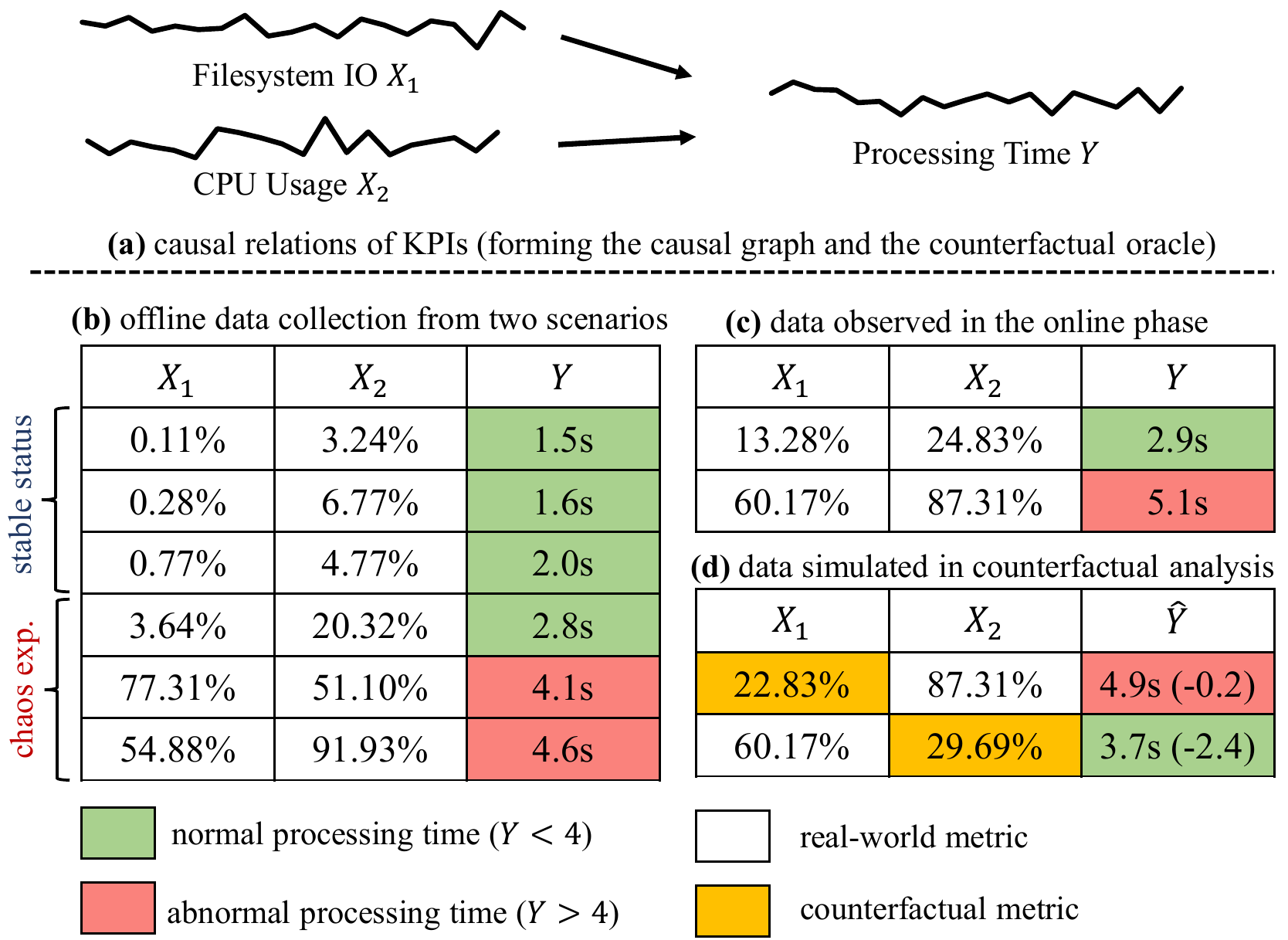}
\caption{Motivating example of performance debugging.}
\label{fig:motivating-example}
\end{figure}

\parh{Usage Scenario.}~Considering \F~\ref{fig:motivating-example}, which
contains three KPIs. Filesystem IO $X_1$ and CPU usage $X_2$, as two causes,
influence the database query processing time $Y$. A developer, Bob, observes a
processing time spike (5.1s) and wonders the root cause of this spike. He
manually checks all performance metrics and identifies that the spike is due to
high CPU usage.

\parh{Ideal Solution.}~As disclosed by vendors~\cite{ma2020diagnosing}, a burst
of performance anomalies may last only a few minutes, whereas human-intensive
diagnoses can take much longer. Causal graphs are an automated and interpretable
solution to this problem, providing informative causal relations among
variables~\cite{gan2021sage}. In this context, Bob would use the causal graph
and a counterfactual oracle, derived from KPI causal relations, to diagnose
performance issues. Bob would first identify all ancestors of the node
indicating processing time $Y$ by traversing the causal graph (see
\F~\ref{fig:motivating-example}(a)). These ancestors represent direct or
indirect causes of performance downgrades. Bob would then submit counterfactual
queries to the oracle to determine whether a particular cause, when changed to a
given extent, could resolve the performance downgrade. The root cause would be
the counterfactual change that fixes the performance downgrade. For example, in
\F~\ref{fig:motivating-example}(d), Bob submits two counterfactual queries
(yellow cells) to the oracle and receives $\hat{Y}$ after such counterfactual
changes. Bob observes that when $X_1$ (filesystem IO) resumes normal operations,
processing time $\hat{Y}$ remains elevated (1st row in
\F~\ref{fig:motivating-example}(d)). When $X_2$ (CPU usage) drops to its mean
value, $\hat{Y}$ returns to normal (2nd row in
\F~\ref{fig:motivating-example}(d)). Thus, Bob attributes this spike to high CPU
usage.


\parh{Challenge.}~Establishing causal graphs and counterfactual oracles is a
long-standing challenge in causality analysis. Existing works use off-the-shelf
causal discovery
algorithms~\cite{chen2014causeinfer,lin2018microscope,jeyakumar2019explainit} or
hand-coded rules~\cite{gan2021sage,wang2021groot} to identify causal graphs from
observational data. A predictive model is then trained on causal relations to
support counterfactual analysis. However, they are insufficient for systematic
performance diagnosis due to \textit{limited
observability}~\cite{spirtes2000causation}. Performance downgrades are rare, so
the causal graph learned from mostly normal states may be biased and unsuitable
for analyzing performance downgrades. Additionally, the predictive model rarely
considers confounders, leading to erroneous counterfactual
predictions~\cite{rosenbaum1983central,hartford2017deep}.

\parh{Our Approach.}~We adopt chaos engineering (CE) to address the
aforementioned challenge. CE is an emerging engineering practice which
extensively injects faults (e.g., network slowdowns) into a system to assess its
reliability. To overcome limited observability, we employ CE to stress various
system events, thus creating sufficient and authentic abnormal data. This
enables us to learn high-quality causal graphs and oracles. These are shown in
the ``chaos exp.'' rows in \F~\ref{fig:motivating-example}(b).


Overall, we present \tool, a CE-enhanced causal analysis framework for
performance debugging. \tool consists of two phases: offline and online. In the
\textit{offline phase}, besides passive observations, CE enables us to actively
collect abnormal data (i.e., those ``chaos exp.'' rows in
\F~\ref{fig:motivating-example}(b)); we can thus learn an accurate and
comprehensive causal graph. Moreover, we use CE to actively mutate chaos
variables (a CE framework typically offers multiple chaos variables, where
mutating each chaos variable can influence hundreds of KPIs), delivering an
accurate estimation of structural equation models (SEMs). We incorporates a set
of design principles and optimizations to overcome challenges (e.g.,
confounders) in the offline phase.

In the \textit{online phase}, when a performance anomaly is observed
(\F~\ref{fig:motivating-example}(c)), we collect the ancestors of the processing
time $Y$ on the causal graph (\F~\ref{fig:motivating-example}(a)) to scope
possible root cause KPIs. Moreover, we use the SEM to issue qualitative
counterfactual queries and obtain accurate root cause analysis, e.g.,
identifying CPU usage $X_2$ as the root cause of the anomaly in
\F~\ref{fig:motivating-example}(d).

\parh{Main Results.}~\tool\ uses an industrial-strength chaos framework,
\cm~\cite{cm}, to establish causal graphs and SEMs with high quality. Evaluation
using synthetic datasets shows that \tool\ offers highly accurate causality
analysis. Moreover, we evaluate \tool\ using MySQL and a distributed database
TiDB~\cite{tidb} on the Kubernetes (K8s)~\cite{k8s} container environments.
Human evaluation show that \tool\ can reliably diagnose performance defects
incurred by various system resources, outperforming existing works with
reasonable cost. In summary, we make the following contributions: 

\begin{itemize}
    \item We for the first time advocate using CE for causality analysis-based
    performance diagnose. Modern CE frameworks, in its ``out-of-the-box''
    manner, principally addresses the low observability issue in causal
    analysis.
    \item We design \tool\ to conduct automated performance anomaly diagnosis
    for complex (distributed) databases. \tool\ incorporates a set of design
    principles and optimizations to enable qualitative root cause identification
    and quantitative counterfactual analysis. We (anonymously) release our
    codebase at~\cite{snapshot} and maintain a documentation at~\cite{doc} to
    help practitioners use \tool.
    \item We evaluate \tool\ on synthetic datasets and real-world (distributed)
    databases. \tool\ shows superior performance over prior works with moderate
    cost.
\end{itemize}

\section{Preliminary}
\label{sec:background}

\subsection{Database Performance Diagnosis}
\label{subsec:perf-debug}

Database developers and users, when encountering performance anomalies, often
aim to identify relevant information for debugging. We classify performance
debugging into two categories based on the available information:  
  
\parh{Blackbox Debugging: Localizing KPIs.}~Database performance downgrades are
often due to abnormal system components like kernel, network, or pod failures in
container clusters~\cite{pod}. In blackbox debugging, users identify root cause
KPIs without accessing database internals. For instance, a developer might want
to determine the cause of an intermittently slow SQL
query~\cite{ma2020diagnosing}, ultimately finding KPIs like high CPU usage or
low disk throughput.
Blackbox debugging is challenging: databases often consist of numerous KPIs
(e.g., our evaluated cases have up to 254 KPIs; see
\S~\ref{sec:implementation}), making it difficult to identify root cause KPIs.
Statistical debugging
(SD)~\cite{jones2005empirical,liblit2005scalable,liu2006statistical} can find
KPIs correlated to anomalies, but correlation does \textit{not} imply causation,
limiting SD's applicability and accuracy. A promising approach is using
\textit{causality analysis}~\cite{gan2021sage, chen2014causeinfer} to establish
causal relations between KPIs, recasting the identification of root cause KPIs
as predicting KPIs with major causation with anomalies based on the causality
graph.  
  
\parh{Whitebox Debugging: Localizing Program Bugs.}~Software bugs may also cause
performance anomalies. This scenario assumes that programmers can monitor
software internals, with typical approaches including software
profiling~\cite{attariyan2012x,zhao2016non},
visualization~\cite{bezemer2015understanding}, and program analysis techniques
like program slicing~\cite{soremekun2021locating}, delta debugging, and
statistical debugging~\cite{jones2005empirical, liblit2005scalable,
liu2006statistical, jin2010instrumentation, zuo2016low}. The end goal is often
to isolate buggy code representing performance bottlenecks.
  
\parh{Focus of This Work.}~Our focus is orthogonal to existing whitebox
debugging tools that aim to find database bugs~\cite{gregg2014systems,
tuya2008controlled, costante2017white}. That is, we perform blackbox debugging
to localize KPIs that result in the performance anomalies. \tool\ provides a
CE-enhanced causal analysis framework for blackbox debugging, and it is agnostic
to specific database implementation. We now introduce preliminaries of causal
analysis and CE.



\subsection{Causality Analysis}
\label{subsec:causality}

\parh{Qualitative Causality Analysis.}~Identifying root causes of performance
anomalies in \F~\ref{fig:motivating-example}, such as abnormal CPU usage leading
to processing time spike, requires a qualitative approach using causality
analysis. This process flags one or multiple KPIs considered as the root causes
of performance defects and requires a causal graph, defined as follows:  
  
\vspace{-2pt}  
\begin{definition}[Causal Graph]  
    A causal graph (Bayesian network) is a directed acyclic graph (DAG):
    $G=(V,E)$. Each node $X$ represents a random variable, and each edge $X\to
    Y$ encodes their cause-effect relationships, with $X$ being a direct cause
    of $Y$. $Pa_G(X)$ denotes the parent nodes of $X$ in $G$.  
\end{definition}  
\vspace{-2pt}  
  
Using causal graphs, identifying anomaly root causes involves determining
cause-effect relations between graph ancestors and descendants. Given an
abnormal KPI $Y$, a common approach backtracks its ancestors and identifies the
most ancestral abnormal KPI $X_i$ as the root cause~\cite{chen2014causeinfer}.

\parh{Quantitative Causality Analysis.} As discussed in
\S~\ref{sec:introduction}, performance anomaly root cause analysis needs a
quantitative perspective of causal relations, typically enabled by
counterfactual analysis using a structural equation model
(SEM)~\cite{peters2017elements} on a qualitative causal graph.  
  
\begin{definition}[SEM]  
    A SEM $M$ consists of:  
    \begin{enumerate}  
        \item Exogenous variables $U$, representing factors outside the model;  
        \item Observed endogenous variables $V$, with each variable $X$ functionally dependent on $U_X\cup Pa_G(X)$, where $U_X\subseteq U$.  
        \item Deterministic functions $f_X \in F$, each $f_X:Pa_G(X)\times U_X\to X$ computing the value of $X$.  
    \end{enumerate}  
\end{definition}  
  
Average treatment effect (ATE)~\cite{holland1986statistics} quantifies the
counterfactual causal effect of treatment $X$ on outcome variable $Y$ in
counterfactual analysis, which answers counterfactual queries:  
\begin{equation}  
    \label{eq:ate2}  
    \text{ATE} = \mathbb{E}[Y \mid do(X=\bm{x}_1)] - \mathbb{E}[Y \mid do(X=\bm{x}_2)]  
\end{equation}  
  
\noindent Here, $do(\cdot)$ represents an intervention on variable $X$.
Intervention sets a variable to a constant value, making it independent of its
parents. The example above asks, "what would $Y$ change if $X$ were $x_1$
instead of $x_2$?" Counterfactual queries like $do(X=x)$ simulate interventions
by removing edges between $X$ and its parent nodes in the SEM, replacing $X$
with a constant $X=x$, while keeping other causal relations
unchanged~\cite{pearl1991theory,balke2022probabilistic}. However, the SEM is
unknown in practice, requiring SEM learning from data and recasting the ATE
causal semantics into a statistical estimand. Computing ATE is challenging when
parent variables interact or are unmeasured in the causal graph. Our solutions
in \S~\ref{subsec:offline} address such challenges and generalize to complex
causal graphs.

\begin{figure}[!ht]
    \centering
    \includegraphics[width=0.85\linewidth]{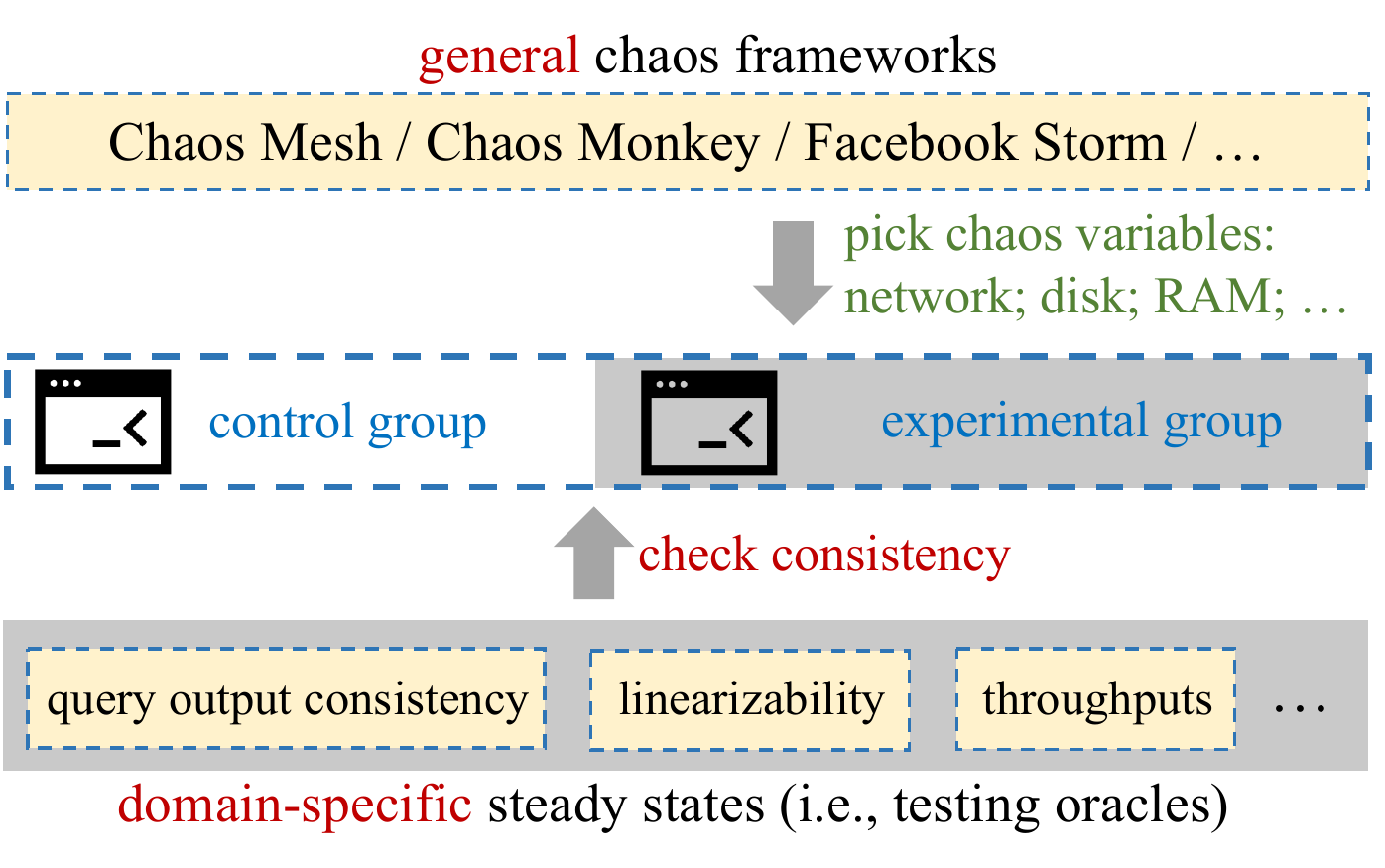}
    \caption{Using CE to aid software testing.}
    \label{fig:chaos}
\end{figure}

\subsection{Chaos Engineering (CE)}
\label{subsec:chaos}

As large-scale, distributed systems evolve, traditional software testing
methods become less effective. CE tests a system's ability to withstand
turbulent conditions in production~\cite{chaos}. In cloud/container
environments, turbulence may include infrastructure, pod, network, and
application failures. \F~\ref{fig:chaos} shows how CE can be used to aid
software testing. In general, CE typically involves three steps: 

\parh{{\Rom{1}}~Defining Steady State.}~CE starts by defining test oracles, or
``steady states,'' as easily measurable outputs of a system that indicate normal
behavior. The key hypothesis is that \textit{a steady state will persist in both
a control group and the experimental group subjected to CE stress.}  
  
Steady states may include throughput and query outputs, considering
domain-specific demands. For example, SQLSmith~\cite{sqlsmith} compares the
outputs of a database (under CE stress) and MySQL. The steady state is defined
as the SQL execution outputs being \textit{consistent} between the two
databases. This setup expands the standard ``differential testing''
procedure~\cite{rigger2020testing,kallas2020diffstream,sotiropoulos2021data},
requiring consistency between the experimental group and reference even under CE
stress. 
  
\parh{{\Rom{2}}~Picking Chaos Variables.}~CE consists of chaos variables, each
representing a critical, low-level factor that may induce failures in
infrastructure, networks, and systems. Modern CE frameworks like \cm~\cite{cm}
are coupled with containerization environments like K8s. offering chaos
variables for various failures in K8s clusters (e.g., container-kill, pod-kill).
Chaos variables are \textit{not} the same as KPIs; there are usually more KPIs
than chaos variables. Mutating each chaos variable (e.g., an IO-related
variable) may affect many KPIs (e.g., average I/O time).  
  
\parh{{\Rom{3}}~Launching CE and Testing.}~After {\Rom{1}} and {\Rom{2}}, CE can
be launched to test the target system, checking for inconsistencies between the
reference group and the target system (experimental group) under CE. Findings
can be used for debugging and error fixing.

\section{Motivation, Related Work and Overview}
\label{sec:motivation}

This section discusses {the} key technical {challenges} of causality
analysis and {reviews} existing works. We then illustrate the synergistic
effect of integrating causality analysis with CE.

\parh{Challenges in Rule-Based Causality Analysis.}~Existing causal-based
performance debugging can be categorized into \textit{rule-based
construction} and \textit{learning-based construction}. However, the
accuracy of estimated causal graphs by both methods {remains}
questionable. First, rule-based construction is highly dependent on expert
knowledge and {is} human-intensive, which may not always {be} correct
with respect to {the} rigorous mathematical properties of causal
relationships~\cite{spirtes2000causation}. In fact, rule-based methods
usually aim at a specific application with limited types of metrics; e.g.,
Sage~\cite{gan2021sage}, one state-of-the-art work, only supports
latency-related KPIs. 

\parh{Challenges in Learning-Based Causality Analysis.}~\response{Rule-based
constructions require domain-specific knowledge (e.g., microservice topological
structures) and can only be applied {to} some specific types of KPIs (e.g.,
Sage is only applicable for latency-related KPIs)~\cite{gan2021sage}. In
contrast, learning-based causality construction {is} generally of higher
applicability than rule-based methods.} Recent
works~\cite{jeyakumar2019explainit,chen2014causeinfer,lin2018microscope} use
learning-based approaches to {create} causal graphs for performance
debugging. This allows {for} more flexible {causal} inference with broader
applications. Nevertheless, in performance debugging, it is challenging to
construct {SEMs} for qualitative/quantitative causality analysis. The key
issue is the \textit{limited observability}. Overall, data collected during
normal database execution {suffers} from selection bias, where abnormal data,
denoting performance anomalies, {is} rare or absent. According to our
observation, a considerable proportion of KPIs (e.g., a KPI denoting failed
queries, known as \texttt{Failed Query OPM} in TiDB~\cite{tidb}) are unchanged
or change negligibly during normal database execution. This hinders learning
accurate causal relations. {In} this regard, existing works often process a
huge amount of logs~\cite{ma2020diagnosing}, thereby subsuming possible
(anomaly) data {with} the best effort.


\begin{figure}[!ht]
  \centering
  \includegraphics[width=0.95\linewidth]{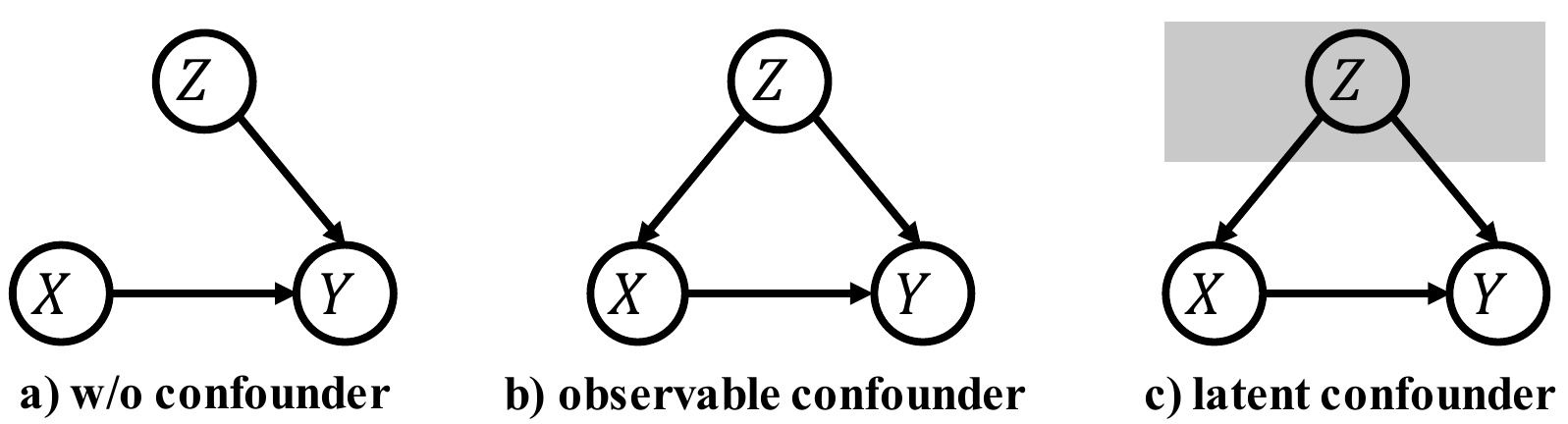}
  \caption{Three typical local structures in a causal graph.}
  \label{fig:local-structure}
  \end{figure}

\parh{Confounders.}~Despite {the} above challenges, existing approaches
neglect one key factor---\textit{confounders}---in causality analysis due to
limited observability. Confounders, either observable or non-observable (called
``latent confounders''), are ubiquitous and hinder causality analysis. For
instance, increasing software input size $Z$ may simultaneously increase
filesystem I/O $X_1$ and processing time $Y$. Therefore, the resulting causality
model that regresses $Y$ on $X_1$ is biased and inaccurate. When $Z$ is observed
(\F~\ref{fig:local-structure}(b)), we may employ {double machine learning
(DML)~\cite{chernozhukov2016double}} to debias. For realistic settings where $Z$
is not observed ($Z$ is a latent confounder, as in
\F~\ref{fig:local-structure}(c)), it is impossible to estimate an unbiased model
from data~\cite{hartford2017deep}.
The state-of-the-art (SOTA) works either assume the absence of
confounders~\cite{gan2021sage} (as in \F~\ref{fig:local-structure}(a)) or assume
that all confounders are observable~\cite{jeyakumar2019explainit}, as in
\F~\ref{fig:local-structure}(b). As shown in \S~\ref{subsec:synthetic},
neglecting observable/latent confounders {impedes} causality analysis
accuracy.

%

\begin{table}[!htbp]
  \centering
  \caption{Comparing existing works and \tool. CE stands for chaos engineering.
  ``B'' and ``W'' in the ``\textbf{Scope}'' column denote blackbox and whitebox views,
  respectively.} 
  \label{tab:comparison}
  \setlength{\tabcolsep}{1.5pt}
\resizebox{\linewidth}{!}{
\begin{tabular}{c|c|c|c|c|c}
    \hline
    \multirow{2}{*}{\textbf{Tool}} & \multirow{2}{*}{\textbf{Scope}} & \textbf{Causal Graph} & \textbf{Root Cause} & \textbf{Counterfactual} & \textbf{General} \\ 
                                                                & & \textbf{Generation}   & \textbf{Analysis}   & \textbf{Analysis?}  & \textbf{KPI?} \\
  \hline
  CauseInfer~\cite{chen2014causeinfer} & B & Learning with data  & Graph Traversal& \XBrush\ & \XBrush\ \\
  \hline
  DBSherlock~\cite{yoon2016dbsherlock} & B & N/A & Actual Cause~\cite{halpern2020causes}& \XBrush\ & \CBrush\  \\
  \hline
  MicroScope~\cite{lin2018microscope} & B & Learning with data  & Graph Traversal& \XBrush\ & \XBrush\ \\
  \hline
  Sieve~\cite{gan2019seer} & B & Granger causality test  & Graph Comparison & \XBrush\ &  \XBrush\ \\
  \hline
  ExplainIt~\cite{jeyakumar2019explainit} & B & Learning with data  & Linear Regression& \XBrush\ & \CBrush\  \\
  \hline
  FluxInfer~\cite{liu2020fluxinfer} & B & N/A  & PageRank~\cite{li1998toward}& \XBrush\ & \CBrush\ \\
  \hline
  iSQUAD~\cite{ma2020diagnosing} & B & N/A  & BCM~\cite{kim2014bayesian}& \XBrush\ & \CBrush\ \\
  \hline
  Sage~\cite{gan2021sage}  & B & Hand-coded rule  & CVAE~\cite{sohn2015learning}& \CBrush\ & \XBrush\ \\
  \hline
  Groot~\cite{wang2021groot}  & W & Hand-coded rule  & PageRank~\cite{li1998toward}& \XBrush\ & \XBrush\ \\
  \hline\hline
  \tool\   & B & CE-enhanced learning  & DML + CE-enabled IV & \CBrush\ & \CBrush\ \\
  \hline
  \end{tabular}
}
\end{table}

\parh{Existing Works.}~\T~\ref{tab:comparison} compares \tool\ with existing
works conceptually and technically. We categorize each work in terms of either
rule-based or learning-based causality analysis. As aforementioned, rule-based
methods are often limited to specific domains and metrics; Sage and Groot
manually defined several rules to constitute causal graphs. As shown in the last
column of \T~\ref{tab:comparison}, they only support limited KPIs (e.g., network
latency-related KPIs) specified in the manual rules. 
Sieve uses Granger causality tests to discover causal relations between {the}
time series. It evaluates if one time series can forecast another, which is not
necessarily the true causality. Most learning-based methods train causal graphs
using offline data collected during normal execution; such passively collected
observations (e.g., system logs) are often biased.

Moreover, given a causal graph, most works rate the contribution of identified
causes using heuristics pertaining to graphical structures, e.g., PageRank-based
solutions add heuristically-designed ``weights'' to graph edges. Such
heuristics-based methods can hardly provide meaningful quantitative relations
between the root cause and performance anomalies. {This prevents} developers
from comprehending how the root cause leads to an anomaly. Methods based on
graph traversal begin with abnormal KPI nodes and backtrack via their ancestors
to identify the root cause. Sage applies predictive models to quantify the
influence of a possible cause. Sage is the only attempt {at} quantitative
counterfactual analysis. Sage, however, only supports quantitative analysis of
network latency-related KPIs, as it primarily focuses on cloud microservices.
Due to limited observability, Sage may be biased with regard to the prevalence
of confounders. \response{DBSherlock uses {an} actual causality
framework~\cite{halpern2005causes_a,halpern2005causes_b} to explain performance
anomalies. However, it employs a simplified actual causality framework that does
not explicitly distinguish cause and effect. As acknowledged in their paper, the
explanation provided by them is different from the actual cause {from a}
causality perspective.}

\parh{\tool.} CE is used for stress testing software systems and in-house
(differential) testing. This paper presents CE as an almost ``out-of-the-box''
option for enhancing causality analysis. CE negatively affects software
performance, with its chaos variables influencing numerous KPIs, providing
abundant training data for learning-based causality analysis. \tool\ supports
general KPIs, not limited to domain-specific instances. Instead of passively
collecting system logs, \tool\ actively mutates chaos variables (influencing
hundreds of KPIs) for comprehensive observability and accurate SEM learning.
This active mutation is referred to as \textbf{active manipulation} in the
paper. Additionally, \tool\ offers quantitative counterfactual analysis with
(latent) confounders and arbitrary KPIs. \tool\ uses double machine learning
(DML) to address observable confounders and instrumental variable (IV) to
overcome latent confounders. This research uses the \textit{domain-general} CE
framework but doesn't require domain-specific ``steady states.'' CE is not
limited to assisting in-house testing, and combined with causality analysis, it
can significantly improve performance debugging of (distributed) databases.  

\begin{figure*}[!ht]
    \centering
    \includegraphics[width=0.90\linewidth]{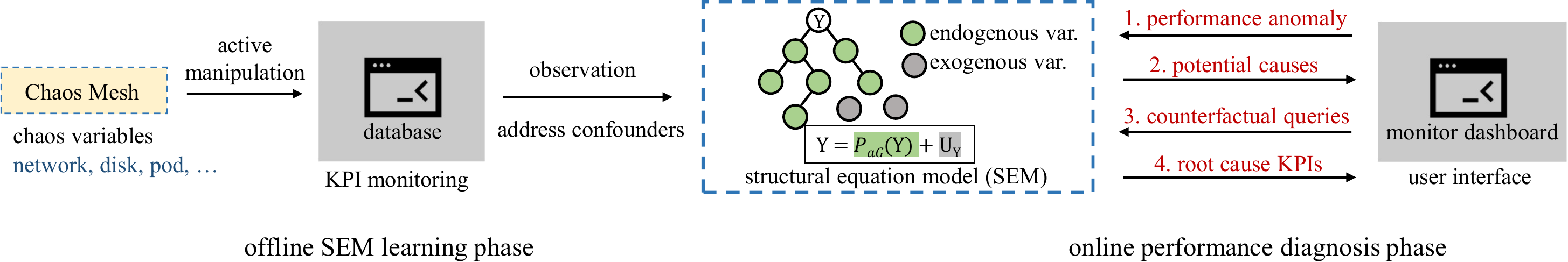}
    \caption{\tool\ overview.}
    \label{fig:overview}
  \end{figure*}

\section{Design of \tool}
\label{sec:overview}


\F~\ref{fig:overview} illustrates \tool's workflow, consisting of offline
learning and online diagnostic phases. The learning steps (structure learning
and parameter learning) are executed before deploying the database to the
production environment, while the online diagnostic phase takes place after its
public release, leading to the classification of these stages as offline and
online, respectively. In the offline phase, \tool\ employs \cm~\cite{cm}, an
industry-standard CE framework integrated into cloud container environments, to
launch CE toward the target databases and gather training data for constructing
SEMs (\S~\ref{subsec:offline}). With active manipulation, CE produces
high-quality SEMs, allowing \tool\ to consider both observable and latent
confounders, as shown in \F~\ref{fig:local-structure}. Using the learned SEM and
a performance anomaly observed during online database execution, \tool\
identifies root cause KPIs. Additionally, \tool\ enables users to pose
counterfactual queries targeting the quantitatively localized root causes, such
as ``\textit{if CPU usage is reduced to 45\%, how would the performance
change?}'' More details can be found in \S~\ref{subsec:online}.

\parh{{User Querying Abnormal KPIs.}}~{We clarify that, during the online
diagnosis phase, \tool\ does not need to decide if its input, a KPI, is
``normal'' or ``abnormal.'' Users can query whatever KPIs they are curious about
and \tool\ will rely on the underlying causal graphs/SEMs to identify root
causes that influence the queried KPIs. Nevertheless, we presume that users
would query \tool\ with abnormal KPIs under performance anomaly cases; this is
obviously the standard usage scenarios of \tool, and it is easy to see that
querying a normal, uninteresting KPI is generally meaningless in our performance
diagnosis context.} 
%


\parh{Application Scope.}~\tool\ debugs real-world database performance
anomalies. Modern databases are complex and susceptible to performance issues.
In our evaluation, we assess \tool\ using synthetic test suites and real-world
databases (MySQL and TiDB), which are also utilized by existing works like
CauseInfer~\cite{chen2014causeinfer}, FluxInfer~\cite{liu2020fluxinfer}, and
ExplainIt~\cite{jeyakumar2019explainit} (see details in \S~\ref{subsec:mysql}).
We deploy MySQL and TiDB in K8s clusters, the de facto system for managing
containerized applications on multiple hosts. Other databases can also be
integrated into K8s and analyzed by \tool.

\subsection{Offline CE-Enhanced SEM Learning Phase}
\label{subsec:offline}

\parh{CE Usage.}~It is a long-standing problem in causal inference to learn
high-quality \rp{SEMs} from observational data. The validity of output SEMs
depends on a set of assumptions. In performance diagnosis, these assumptions are
frequently violated, posing \rp{obstacles} to subsequent causality analysis. We
employ CE in SEM learning: In \response{\Rom{1}} and \response{\Rom{2}}, we
explain how CE enables accurate and comprehensive SEM learning. In
\response{\Rom{3}}, we detail the operations for generating training data using
CE. 

\sparh{\response{\Rom{1}} Improving Observability on Normal
Data.}~According to the literature~\cite{spirtes2000causation},
\textit{faithfulness} guarantees d-separation on causal graphs. This
assumption allows for refuting causal relations without correlations.
However, insufficient data can lead to a lack of correlations on causally
related data. Many KPIs remain unchanged or change little in daily states,
making it difficult to collect sufficient observations to the underlying
system and establish causal relations. \tool\ uses \cm\ for ``active
manipulation'' during causal learning: \cm\ enables mutating chaos
variables to improve observability on value ranges and KPI causality. For
example, mutating the ``network loss probability'' affects network-related
KPIs, like ``network duration of request.'' CE explores a larger KPI value
space and effectively uncovers infrequently occurring KPI states, thereby
building a more accurate causal graph.

\sparh{\response{\Rom{2}} Improving Observability on Anomalies.}~CE
allows for triggering anomalies offline by heavily mutating certain chaos
variables. By inspecting real performance anomalies, we find that many
basic performance anomalies (e.g., ``I/O Saturation'') are outcomes of the
I/O-concerned chaos variable. Moreover, some complex performance anomalies
(e.g., ``Workload Spike'') can be simulated by combining chaos variables. This
way, we not only augment the comprehensiveness of observed normal data (as in
\response{\Rom{1}}), but also largely enrich the knowledge of anomaly data. This
would effectively alleviate the OOD (out-of-distribution) challenge faced by
previous works\response{, which are presented in \T~\ref{tab:comparison}}, since
online anomaly data was rarely seen in offline training data collected by
previous research.

\sparh{\response{\Rom{3}} Collecting Training Data.}~To collect offline
training data, in addition to the constant queries emitted by workload
simulators (such as the TPC-C benchmark~\cite{tpcc}), we also use a set of
chaos variables to stress the target databases. Among the chaos variables
included in our experiments (see details in \S~\ref{sec:implementation}),
some, when enabled, directly inject faults into the target databases.
When generating training data, we enable each of them and record how KPIs
change. Other chaos variables are \textit{configurable}, allowing us to
mutate their values to determine the extent of their impact on databases.
When generating training data, for each of them, we divide its valid input
range equally into three to five thresholds (depending on the sensitivity of
this chaos variable; more sensitive variables are mutated five times
and vice versa). We mutate each chaos variable using all of its thresholds,
and we collect the KPI changes for each iteration. KPIs are initially used
to learn the structure of the causal graph, and then chaos variables and
KPIs are used together to learn the SEM's parameters (see details in the
following paragraphs).

\parh{Causal Graph Structure Learning.}~Once \rp{the} offline training data
\rp{is gathered}, we first learn causal graphs. Causal graph structure
learning generates a DAG, where each node denotes one KPI and each edge
represents \rp{the causal} relations of a KPI pair. This enables qualitative
\rp{causality} analysis, such that given a performance anomaly monitored
during the online phase (denoting a node on the DAG), we collect its
ancestors to form the potential root causes (see details in
\S~\ref{subsec:online}). Previous work like
CauseInfer~\cite{chen2014causeinfer} has relied heavily on constraint-based
algorithms (e.g., the PC algorithm~\cite{spirtes2000causation}), which
returns an incomplete causal graph with undirected edges (CPDAG) and
arbitrarily \rp{assigns} directions to undirected edges. In \tool, we aim to
harness the power of score-based algorithms to learn a causal graph that
maximizes a predefined score over \rp{the} observational data, with all
edges in the resulting graph being directed. In particular, we employ a
two-stage technique based on BLIP~\cite{scanagatta2015learning} to learn
causal graphs. To begin \rp{with}, for each variable, we identify its
possible parent sets with local scores. \rp{We then} use a global structure
optimization algorithm to identify the causal graph that maximizes the
global score, which is computed using cached local scores in the first
stage. According to the literature~\cite{ding2020reliable}, BLIP produces
empirically much better results than the PC
algorithm~\cite{spirtes2000causation} employed in previous works.
Nevertheless, \tool\ is not bounded \rp{by} BLIP; we use it as it is one
recent work that offers off-the-shelf implementation with good engineering
quality. Users can replace BLIP with other algorithms whenever
needed~\cite{spirtes2000causation,ding2020reliable}. 


\parh{Causal Graph Parameter Learning.}~Given \rp{that} the DAG representing
causal graph structures, we further conduct parameter learning to assign
relationships toward each edge of the DAG. This step enables quantitative
counterfactual analysis. Considering the sample case in
\F~\ref{fig:motivating-example}, given filesystem I/O $X_1=x_1$ and CPU usage
$X_2=x_2$, quantitative causality analysis infers the consequent processing time
$Y$. For this simple case, we can directly train a predictive model that
predicts $Y$ given $X_1$ and $X_2$. \response{\rp{This} trained model can be
directly used to conduct counterfactual analysis, i.e., computing the ATE (refer
to \E~\ref{eq:ate2}) of $X_1$ or $X_2$ on $Y$.}

\begin{figure}[!ht]
    \centering
    \includegraphics[width=0.60\linewidth]{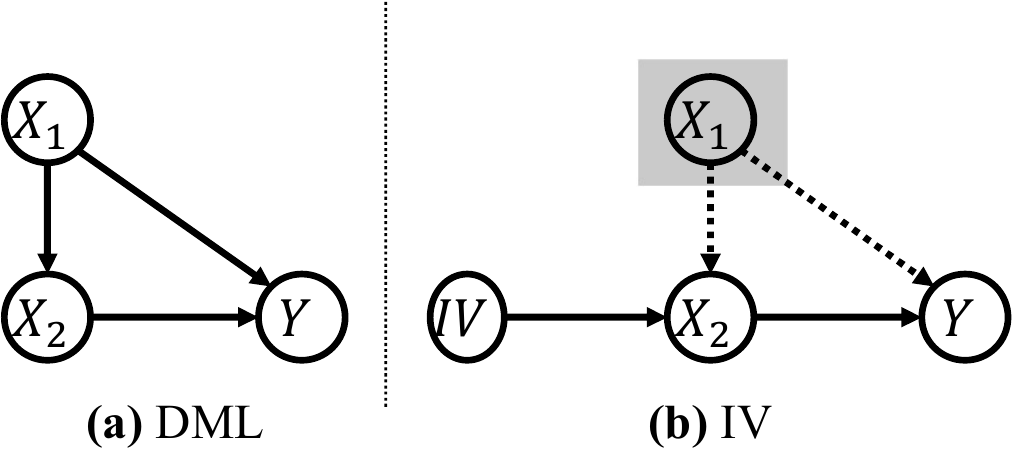}
    \caption{\response{Counterfactual prediction using DML or IV.}}
    \label{fig:iv}
\end{figure}

However, such methods are \textit{not} suitable when $X_1$ and $X_2$ have a
causal relationship ($X_1$ forms a confounder). For instance, heavy
filesystem I/O itself may result in poor CPU usage, as processes may be
blocked while waiting for I/O. In this case, $X_1$ is an observable
confounder of $X_2$. We start to analyze a potential solution by assuming
linear causal relations and will extend it to general cases later. In this
setting, $X_1$ is an exogenous variable, while $X_2$ and $Y$ are endogenous
variables (\F~\ref{fig:iv}(a)). Suppose $X_2=2X_1+b_1$ and
$Y=X_1+2X_2+b_2$, where $b_1$ and $b_2$ are constants, and the zero-mean
noise (or disturbance) are omitted for simplicity. As a standard solution,
it is feasible to use a linear model to fit the data. However, the
parameters of the learned model may be biased, which undermines subsequent
counterfactual analysis (i.e., computing ATE). For instance,
$Y=5X_1+2b_1+b2$ also forms a valid solution, when predicting $Y$ given
$X_1$ and $X_2$. However, it fails to enable counterfactual analysis, as
$X_2$'s effect on $Y$ is not captured. 
It is worth noting that this issue could impose notable challenges to any
machine learning models, as long as they directly use $X_1,X_2$ to regress
$Y$. We now introduce how \tool\ handles observable and latent confounders.

\parh{Observable Confounders.}~We use double machine learning
(DML)~\cite{chernozhukov2016double} to handle observable confounders.
Continuing the previous example (\F~\ref{fig:iv}(a)),
DML predicts $\hat{Y}$ from $X_1$ and $\hat{X_2}$, respectively. It then
estimates the relationship between the residuals by training a model to
predict $Y-\hat{Y}$ using $X_2-\hat{X_2}$ as input. Finally, this model
estimates the true causal effect of $X_2$ on $Y$ (i.e., ATE) and can be
extended to arbitrary functions, including deep learning models for
non-linear causal relations.


\parh{Latent Confounders.}~The above technique is still based on the premise
that all confounders (i.e., $X_1$ in our example) are observable. It cannot be
extended to a more complex setting with latent confounders, which pervasively
exist in real-world settings. Overall, the presence of latent confounders makes
counterfactual predictions technically challenging, if not impossible. This is
because we cannot disentangle the effect of $X_1$ from the causal relationship
between $X_2$ and $Y$. We note that none of \rp{the} existing works reviewed in
\T~\ref{tab:comparison} ever consider latent confounders, or they simply assume
the absence of latent confounders.


A commonly-used tactic, namely instrumental variable (IV), addresses
learning a regression function under an  ``interventional
distribution''~\cite{pearl2009causality}. In \F~\ref{fig:iv}(b), IV serves
as the cause of treatment ($X_2$), and its effect on the outcome ($Y$) is
propagated with $X_2$. A valid IV should have only one causal effect on $Y$
mediated by $X_2$, meaning the causal relationship between $IV$ and $Y$
must be indirect.  

It is unclear how prior works can leverage IV, given \rp{that} they primarily
collect system logs in a \textit{passive} manner. In contrast, \tool\
\textit{novelly} uses CE to form the IV, which has a direct influence on the CPU
usage $X_2$ and an indirect effect on processing time $Y$ via $X_2$. We clarify
that CE is proper to form IV: CE typically mutates low-level chaos variables
(e.g., CPU workload, network delay time) by injecting failures. Thus, it is
reasonable to assume the causal relationships between CE (through the mutated
chaos variables) and the influenced ``high-level'' KPIs are indirect. This
characteristic makes these chaos variables comply \rp{with} the above
prerequisite, providing a distinct opportunity to smoothly treat chaos variables
as instrumental variables and apply \rp{an} IV framework accordingly. 
Without loss of generality, we use \F~\ref{fig:iv}(b) to illustrate the
standard procedure of IV. IV first fits a model to predict $X_2$ from $IV$
\rp{in order} to get the predicted $\hat{X_2}$. Then, it fits another model
to predict $Y$ from $IV$ \rp{in order} to get the predicted $\hat{Y}$.
Afterwards, IV fits a model to predict $\hat{Y}$ from $\hat{X_2}$. This
trained model can be used to estimate the true causal effect of $X_2$ on
$Y$ (i.e., ATE). Similarly, this pipeline is extensible to arbitrary
functions. To ensure generalizability, \tool\ uses
DeepIV~\cite{hartford2017deep}, a popular IV framework that allows
counterfactual analysis with deep learning models. This helps approximate
nonlinear function forms.

\begin{algorithm}[!htbp]
\scriptsize
\caption{Root Cause Analysis (RCA)}
\label{alg:rca}
\KwIn{Observed KPIs $\bm{x}=(\bm{x}_1,\cdots,\bm{x}_n)$, 
KPI of Interest $Y$, SEM $M$}
\KwOut{Ordered Potential Root Causes $\bm{X}=\{X_i,\cdots\}$}
$y\leftarrow \bm{x}_Y$; // observed KPI of interest\\
$\bm{X}_C=\{X_a\mid X_a \text{~is an ancestor of }Y \text{~on } M\}$; //initialize candidates\\
\ForEach{$X_a\in \bm{X}_C$}{
  $\hat{y}\leftarrow\mathbb{E}[ Y\mid \overbrace{X_j=\bm{x}_j}^{\forall j\neq a},do(X_a=\mathbb{E}\left[X_a\right])]$\;
  $s_a\leftarrow \texttt{PDF}_{Y}(\hat{y}) - \texttt{PDF}_{Y}(y)$\; 
  assign $s_a$ as $X_a$'s blame\;
}
$\bm{X}\leftarrow\{X_a\mid X_a \in \bm{X}_C, s_c>0\}$\;
rank $\bm{X}$ by blames\;
\Return{$X$}
\end{algorithm}

\subsection{Online Root Cause Analysis}
\label{subsec:online}

\parh{Workflow.}~\A~\ref{alg:rca} shows the root cause analysis process for
performance anomalies during the online phase. $\bm{x}$ is a vector of KPIs
monitored by the database, $Y$ represents the user-specified KPI of interest
(e.g., query processing time), and $M$ is the SEM from the offline phase.

Assuming $Y$ is the performance anomaly KPI, indicating high database load,
\A~\ref{alg:rca} extracts the current $Y=y$ observation from $\bm{x}$ (line 1).
Next, it traverses the causal graph $M$ to find $Y$'s ancestors as potential
root causes (line 2). For each candidate, the algorithm calculates its blame
(lines 3--7) by predicting a counterfactual expectation under an
intervention\footnote{Interventions in our counterfactual analysis are
simulated, distinct from manual causal graph intervention. See
\S~\ref{subsec:causality} for more details.} on the current $X_a$ (line 4). It
estimates ``how $Y$ would be if $X_a$ were normal,'' where ``normal'' means
$X_a=\mathbb{E}\left[X_a\right]$ (line 4). The blame score of $X_a$ is computed
by comparing the probability density of $y$ and the estimated counterfactual
value $\hat{y}$ (lines 5--6), where $\texttt{PDF}_{Y}$ is the probability
density function of $Y$. By ranking causes by blame, \A~\ref{alg:rca} directs
users' attention to the most influential root cause KPIs. The following sections
detail counterfactual predictions and probability densities.

\begin{algorithm}[!htbp]
\scriptsize
\caption{Counterfactual Prediction}
\label{alg:cp}
\KwIn{Observed KPIs $\bm{x}=(\bm{x}_1,\cdots,\bm{x}_n)$, 
KPI of Interest $Y$, Candidate Cause $X_a$, SEM $M$}
\KwOut{Counterfactual Prediction $\hat{y}$}
$\hat{\bm{x}}\leftarrow \bm{x}$\;
$\hat{\bm{x}_a}\leftarrow \mathbb{E}[X_a]$\;
$\bm{X}_D=\{X_d\mid X_d \text{~is a descendant of }X_a \text{~on } M\}$\;
sort $\bm{X}_D$ in the topological order on $M$\;
\ForEach{$X_d\in \bm{X}_D$}{
  Total\_TE $\leftarrow \sum_{X_p\in Pa(X_d)} \left(\mathbb{E}[X_d\mid do(X_p=\hat{\bm{x}_p})] - \mathbb{E}[X_d\mid do(X_p=\bm{x}_p)]\right)$\;
  $\hat{\bm{x}_d} \leftarrow \bm{x}_d + $Total\_TE\;
  \lIf{$X_d$ is $Y$}{break}
}
\Return{$\hat{\bm{x}_Y}$}
\end{algorithm}

\parh{Counterfacutal Prediction.}~Performing counterfactual predictions is
difficult, given the complexity of causal structures. As noted on line 4 of
\response{\A~\ref{alg:rca}}, we proceed \rp{with} the counterfactual predictions
as follows:
\begin{equation}
  \footnotesize
  \label{eq:counterfactual}
  \hat{y}=\mathbb{E}[ Y\mid \overbrace{X_j=\bm{x}_j}^{\forall j\neq a},do(X_a=\mathbb{E}\left[X_a\right])]
\end{equation}

\noindent which estimates the conditional expectation of $Y$ given the currently
observed KPIs $\bm{x}$ and an intervention on one KPI, $X_a$ while keeping the
remaining KPIs ($j\neq a$) unchanged. When $X_a$ is a simple direct cause of
$Y$, estimating \E~\ref{eq:counterfactual} is straightforward: we can simply add
the average treatment effect (ATE\response{; see \E~\ref{eq:ate2}}) of 
$X_a$ on $Y$ to the currently observed $y$. Formally,
\begin{equation}
  \footnotesize
  \hat{y}=y + (\mathbb{E}[Y \mid do(X_a=\mathbb{E}[X_a])] - \mathbb{E}[Y \mid do(X_a=\bm{x}_a)])
\end{equation}

\noindent where the second operand (in the parentheses) of the addition
represents the ATE of $X_a$ on $Y$, which can be further computed using the SEM
learned in the offline phase. 

%
However, in general cases where $X_a$ is an ancestor of $Y$, counterfactual
changes on $X_a$ may influence other KPIs, whose changes may propagate to $Y$ as
well. To systematically model the effect of $X_a$ on $Y$, we recursively update
all $X_a$'s descendants on the causal graph with respect to the counterfactual
change and estimate $\hat{y}$. 
\A~\ref{alg:cp} outlines the procedure. We use $\hat{\bm{x}}$ to maintain the
counterfactual values of KPIs (line 1) and update $\hat{\bm{x}_a}$ to its mean
value (line 2). With $do(X_a=\mathbb{E}[X_a])$, only the descendants of $X_a$
will be modified. Therefore, we estimate each of its descendants in topological
order (lines 3--9) such that a descendant is updated when only all its ancestors
have been updated to the counterfactual values. For each descendant $X_d$, we
compute the total treatment effect (Total\_TE) as the sum of its parents'
treatment effects on $X_d$ (line 6) and derive the counterfactual $X_d$
accordingly (line 7). When $Y$ is updated, we terminate the procedure and return
$\hat{\bm{x}_Y}$ as the counterfactual prediction of $Y$ given the intervention
of $do(X_a=\mathbb{E}[X_a])$.

\parh{Probabilistic Modeling.}~Computing $\texttt{PDF}_{Y}(y)$ is challenging
due to the unknown distribution of $X_k$. Assuming a prior distribution family
and estimating parameters from observations makes computing
$\texttt{PDF}_{Y}(y)$ feasible. However, neither Gaussian nor beta distributions
accurately estimate real-world KPIs. For instance, when dealing with multiple
modals, these assumptions fall short. We use Gaussian
KDE~\cite{rosenblatt1956remarks,parzen1962estimation} for non-parametric
approximation, consistently yielding good performance across KPIs.  

\section{Implementation}
\label{sec:implementation}

We implement \tool\ in 2K lines of Python code and \cm\ (ver. 2.1.3) with 3K
lines of yaml configuration. \tool\ works with K8s, making it compatible with
various DBMSs (e.g., MySQL, PostgreSQL, TiDB) and adaptable for performance
debugging in different scenarios. We use K8s due to its integration with \cm\
and it enables analyzing distributed cloud databases. See the chaos variables
list at~\cite{chaosvar}.

\parh{Databases.}~We apply \tool\ to MySQL and TiDB. We follow prior
works~\cite{yoon2016dbsherlock} and deploy MySQL with a single instance on K8s.
For TiDB, we use a recommended test cluster with one \texttt{PD} pod, one
\texttt{TiDB} pod, and three \texttt{TiKV} pods on a 3-node K8s
cluster~\cite{tidbConfig}. \texttt{PD}, \texttt{TiDB}, and \texttt{TiKV} serve
as manager, interface, and storage components, respectively. In K8s, a ``pod''
is an application's minimal unit. The 3-node cluster consists of a control-plane
and two worker nodes.

\parh{KPI Monitoring.}~\tool\ uses Grafana~\cite{grafana} and
Prometheus~\cite{prometheus} for KPI monitoring in K8s. The details of our
employed KPIs can be found in~\cite{mysqlkpi,tidbkpi}. \tool\ records each KPI's
value every one second.

\parh{Offline Training.}~\S~\ref{subsec:offline} explains that \tool\ extends
the codebase of BLIP~\cite{scanagatta2015learning} to learn causal graphs. We
adopt BLIP due to its SOTA performance and high engineering quality. We extend
its codebase with 112 extra lines of Python code. \tool\ is orthogonal to
particular causal graph learning algorithms. We leave it as one future work to
explore leveraging other SOTA algorithms like REAL~\cite{ding2020reliable} and
ML4C~\cite{dai2021ml4c}. We employ EconML~\cite{econml} to perform double
machine learning and DeepIV~\cite{hartford2017deep} to estimate the parameters
of causal relationships, where random forest models and neural networks (Multi
Layer Perception) are jointly employed for regression. We present model details
in \rp{the} \sm.


\section{Evaluation}
\label{sec:evaluation}

We conduct extensive evaluations on MySQL in
\S~\ref{subsec:mysql} to demonstrate \tool's effectiveness in practice. To
study its scalability, we evaluate \tool\ on a large distributed DBMS,
TiDB, in \S~\ref{subsec:tidb}. In addition, we compare \tool's
counterfactual analysis with machine learning models on synthetic data in
\S~\ref{subsec:synthetic}.

\parh{Processing Time.}~For data collection, we run the database for six hours
without chaos mesh experiments. Then, we execute a 12-hour chaos mesh workflow
on MySQL and a 20-hour workflow on TiDB. Each workflow comprises multiple chaos
experiments, with a 10-minute suspend period between them. Thus, we gather 18
hours of MySQL data and 26 hours of TiDB data. While chaos mesh introduces
non-trivial overhead, prior
works~\cite{yoon2016dbsherlock,liu2020fluxinfer,gan2021sage} often collect data
over longer periods.
  
The offline training phase is a one-shot effort. Users should be more concerned
with potential online overhead: \tool\ incurs negligible online overhead, as all
online analysis tasks are rapid. For MySQL, it takes about 50 minutes for causal
graph structure offline learning and 8 minutes for parameter learning. Analyzing
one performance anomaly (quantitative and qualitative aspects) takes less than a
minute. For TiDB, \tool\ requires about 1.3 hours for causal graph learning and
40 minutes for parameter learning. In the online inference phase, each anomaly
analysis takes 3 to 5 minutes, depending on the number of involved KPIs.
Evaluating synthetic datasets takes several seconds.  


\subsection{Evaluation on Effectiveness}
\label{subsec:mysql}

\begin{table}[t]
  \centering
  
  \caption{Performance anomalies on MySQL.}
  \resizebox{0.95\linewidth}{!}{
    \begin{tabular}{c|c||c|c}
      \hline
      \textbf{Component}        & \textbf{Anomaly}   &\textbf{Component}        & \textbf{Anomaly}\\
      \hline
      Client                    & Workload Spike    & \multirow{3}{*}{I/O}  & I/O Saturation     \\\cline{1-2}\cline{4-4}
      \multirow{4}{*}{Database} & Database Backup   &  & I/O Latency         \\\cline{2-2}\cline{4-4}
                                & Database Restore  & & I/O Fault  \\\cline{2-2}\cline{3-4}
                                & Flush Log         & \multirow{2}{*}{Network} &  Network Delay  \\\cline{2-2}\cline{4-4}
                                & Flush Table     & & Network Partition
      \\\hline
      Memory                    & Memory Stress       & CPU                       & CPU Stress                        \\\hline
    \end{tabular}
  }
  \label{tab:mysql-anomaly}
\end{table}


\parh{Environment and Data Collection.}~We set up a MySQL instance on K8s,
and execute TPC-C queries using BenchBase~\cite{DifallahPCC13} for a
duration of 18 hours. To collect relevant KPIs, we utilize
mysql-exporter~\cite{mysql-exporter}, covering a total of 79 KPIs. During
the online phase, we simulate performance anomalies by following the
configuration in DBSherlock~\cite{yoon2016dbsherlock}, as shown in
\T~\ref{tab:mysql-anomaly}. These anomalies correspond to various key
components in MySQL. In total, we create 13 anomaly instances that could
potentially have negative effects on user-focused KPIs (e.g.,
Query\_Duration).


\parh{Baselines.}~We re-implement CauseInfer~\cite{chen2014causeinfer},
FluxInfer~\cite{liu2020fluxinfer}, and ExplainIt~\cite{jeyakumar2019explainit}
as baseline methods, as reviewed in \T~\ref{tab:comparison}. For
CauseInfer~\cite{chen2014causeinfer}, we implement a variant (CauseInfer+CE) by
augmenting its causal discovery process with our CE module. We exclude methods
without general KPI support and omit DBSherlock~\cite{yoon2016dbsherlock} and
iSQUAD~\cite{ma2020diagnosing}, as they require labeling normal and abnormal
regions during the offline phase.

\parh{Human Evaluation Setup.}~\tool\ and three baseline methods provide ranked
KPI lists, and we collect the top-5 KPIs for each anomaly experiment. We invite
ten experts to select relevant KPIs from those recommended by at least one
method. After briefing them on the database performance debugging scenario and
KPI meanings, they choose KPIs closely related to each anomaly's root cause.
Each KPI is assigned a score as $\frac{\text{\# vote}}{\text{\# participant}}$
for a given anomaly, with a higher score indicating a higher expert consensus.
Treating expert-selected KPIs as ground truth, we evaluate the methods'
effectiveness in identifying root causes. The average completion time is 45
minutes, with details in the supplementary material.


\parh{Metric.}~Using the scores from human evaluation, we consider the following
metrics for comparison. We compute the average score (AS) of the top-1/3/5 KPIs
suggested by each method. Additionally, we employ MAP@R (Mean Average Precision)
and NDCG (Normalized Discounted Cumulative Gain), two widely-used metrics in
recommendation systems, to evaluate the accuracy of the ranked KPI lists in
relation to user preferences.

\begin{table}[t]
   \scriptsize
  \caption{\response{Comparison of different methods. \#Anomaly denotes the number of
    anomalies in this component. We highlight the \colorbox{best}{best} and
    \colorbox{second}{second best} methods in each component.}}
  \label{tab:full-comparison}
  \begin{center}
    \resizebox{1.0\linewidth}{!}{\begin{tabular}{
        @{\hspace{1pt}}c@{\hspace{2.0pt}}
        @{\hspace{1.5pt}}c@{\hspace{1.5pt}}
        @{\hspace{1.5pt}}c@{\hspace{1.5pt}}
        @{\hspace{1.5pt}}c@{\hspace{1.5pt}}
        @{\hspace{1.5pt}}c@{\hspace{1.5pt}}
        @{\hspace{1.5pt}}c@{\hspace{1.5pt}}
        @{\hspace{1.5pt}}c@{\hspace{1pt}}
        }
        \toprule
        \textbf{Component}   & \multirow{2}{*}{\textbf{Metric}} & \multirow{2}{*}{\textbf{\tool}} & \multirow{2}{*}{\textbf{CauseInfer}} & \multirow{2}{*}{\textbf{CauseInfer+CE}} & \multirow{2}{*}{\textbf{FluxInfer}} & \multirow{2}{*}{\textbf{ExplainIt}} \\
        \textbf{(\#Anomaly)} &                                  &                                 &                                      &                                         &                                     &                                     \\
        \midrule
        \multirow{5}{*}{\makecell{Client                                                                                                                                                                                                                       \\(1)}}
                             & Top-1 AS                         & \colorbox{second}{0.70}         & 0.00                                 & 0.00                                    & \colorbox{best}{0.90}               & 0.00                                \\
                             & Top-3 AS                         & \colorbox{second}{0.73}         & 0.00                                 & 0.00                                    & \colorbox{best}{0.90}               & 0.20                                \\
                             & Top-5 AS                         & \colorbox{best}{0.80}           & 0.00                                 & 0.00                                    & \colorbox{second}{0.70}             & 0.12                                \\
                             & MAP@R                            & \colorbox{second}{0.36}         & 0.00                                 & 0.00                                    & \colorbox{best}{0.41}               & 0.02                                \\
                             & NDCG                             & \colorbox{best}{0.82}           & 0.00                                 & 0.00                                    & \colorbox{second}{0.81}             & 0.11                                \\
        \cmidrule(r){1-7}
        \multirow{5}{*}{\makecell{Database                                                                                                                                                                                                                     \\(4)}}
                             & Top-1 AS                         & \colorbox{best}{0.40}           & 0.00                                 & \colorbox{second}{0.18}                 & \colorbox{best}{0.40}               & 0.00                                \\
                             & Top-3 AS                         & \colorbox{best}{0.49}           & 0.14                                 & 0.32                                    & \colorbox{second}{0.36}             & 0.05                                \\
                             & Top-5 AS                         & \colorbox{best}{0.56}           & 0.09                                 & 0.28                                    & \colorbox{second}{0.40}             & 0.09                                \\
                             & MAP@R                            & \colorbox{best}{0.17}           & 0.02                                 & 0.10                                    & \colorbox{second}{0.13}             & 0.01                                \\
                             & NDCG                             & \colorbox{best}{0.65}           & 0.11                                 & 0.38                                    & \colorbox{second}{0.45}             & 0.08                                \\
        \cmidrule(r){1-7}
        \multirow{5}{*}{\makecell{I/O                                                                                                                                                                                                                          \\(4)}}
                             & Top-1 AS                         & \colorbox{best}{0.93}           & \colorbox{second}{0.40}              & 0.25                                    & 0.23                                & 0.03                                \\
                             & Top-3 AS                         & \colorbox{best}{0.81}           & 0.28                                 & \colorbox{second}{0.43}                 & 0.15                                & 0.08                                \\
                             & Top-5 AS                         & \colorbox{best}{0.76}           & 0.18                                 & \colorbox{second}{0.37}                 & 0.25                                & 0.18                                \\
                             & MAP@R                            & \colorbox{best}{0.30}           & 0.08                                 & \colorbox{second}{0.12}                 & 0.08                                & 0.03                                \\
                             & NDCG                             & \colorbox{best}{0.91}           & 0.27                                 & \colorbox{second}{0.42}                 & 0.28                                & 0.16                                \\
        \cmidrule(r){1-7}
        \multirow{5}{*}{\makecell{Network                                                                                                                                                                                                                      \\(2)}}
                             & Top-1 AS                         & \colorbox{best}{0.90}           & \colorbox{second}{0.35}              & 0.20                                    & 0.20                                & 0.00                                \\
                             & Top-3 AS                         & \colorbox{best}{0.90}           & 0.25                                 & \colorbox{second}{0.37}                 & 0.12                                & 0.08                                \\
                             & Top-5 AS                         & \colorbox{best}{0.90}           & 0.15                                 & \colorbox{second}{0.32}                 & 0.24                                & 0.15                                \\
                             & MAP@R                            & \colorbox{best}{0.34}           & 0.07                                 & \colorbox{second}{0.10}                 & 0.08                                & 0.03                                \\
                             & NDCG                             & \colorbox{best}{0.99}           & 0.22                                 & \colorbox{second}{0.33}                 & 0.25                                & 0.13                                \\
        \cmidrule(r){1-7}
        \multirow{5}{*}{\makecell{Memory                                                                                                                                                                                                                       \\(1)}}
                             & Top-1 AS                         & \colorbox{best}{1.00}           & \colorbox{best}{1.00}                & \colorbox{second}{0.70}                 & \colorbox{second}{0.70}             & 0.00                                \\
                             & Top-3 AS                         & \colorbox{best}{1.00}           & 0.33                                 & \colorbox{second}{0.80}                 & 0.23                                & 0.00                                \\
                             & Top-5 AS                         & \colorbox{best}{1.00}           & 0.33                                 & \colorbox{second}{0.70}                 & 0.36                                & 0.20                                \\
                             & MAP@R                            & 0.17                            & 0.08                                 & \colorbox{best}{0.63}                   & \colorbox{second}{0.25}             & 0.03                                \\
                             & NDCG                             & \colorbox{best}{1.00}           & 0.32                                 & \colorbox{second}{0.92}                 & 0.51                                & 0.17                                \\
        \cmidrule(r){1-7}
        \multirow{5}{*}{\makecell{CPU                                                                                                                                                                                                                          \\(1)}}
                             & Top-1 AS                         & \colorbox{best}{0.60}           & 0.00                                 & 0.00                                    & \colorbox{second}{0.30}             & 0.00                                \\
                             & Top-3 AS                         & \colorbox{second}{0.60}         & 0.00                                 & 0.00                                    & \colorbox{best}{0.77}               & 0.00                                \\
                             & Top-5 AS                         & \colorbox{second}{0.60}         & 0.00                                 & 0.00                                    & \colorbox{best}{0.66}               & 0.12                                \\
                             & MAP@R                            & \colorbox{second}{0.12}         & 0.00                                 & 0.00                                    & \colorbox{best}{0.51}               & 0.02                                \\
                             & NDCG                             & \colorbox{second}{0.60}         & 0.00                                 & 0.00                                    & \colorbox{best}{0.74}               & 0.09                                \\
        \cmidrule(r){1-7}
        \multirow{5}{*}{\makecell{Overall                                                                                                                                                                                                                      \\(13)}}
                             & Top-1 AS                         & \colorbox{best}{0.72}           & 0.25                                 & 0.22                                    & \colorbox{second}{0.37}             & 0.01                                \\
                             & Top-3 AS                         & \colorbox{best}{0.72}           & 0.19                                 & 0.35                                    & \colorbox{second}{0.32}             & 0.07                                \\
                             & Top-5 AS                         & \colorbox{best}{0.73}           & 0.13                                 & 0.30                                    & \colorbox{second}{0.37}             & 0.14                                \\
                             & MAP@R                            & \colorbox{best}{0.25}           & 0.05                                 & 0.12                                    & \colorbox{second}{0.17}             & 0.02                                \\
                             & NDCG                             & \colorbox{best}{0.80}           & 0.18                                 & 0.37                                    & \colorbox{second}{0.42}             & 0.12                                \\
        \cmidrule(r){1-7}
      \end{tabular}}
  \end{center}
\end{table}

\parh{Result Overview.}~We report the results of Top-1/3/5 Average Score (AS),
MAP@R and NDCG in \T~\ref{tab:full-comparison}. We observe that \tool\
substantially outperforms existing methods for nearly all settings. For
instance, its overall NDCG is \response{90.5\%} higher than the second
best method, FluxInfer.
It {also} provides the best root causes for \response{five out of seven}
anomaly components compared to {the} baseline methods. In short, we
consider \tool\ {to be} sufficient {enough} to produce reasonable KPIs
to assist developers in identifying the root cause. Empirical results also
show that CE offers a general augmentation toward causality-based
approaches. As {shown} in \T~\ref{tab:full-comparison}, after being
augmented with CE, CauseInfer manifests much better performance in most
settings (\response{106\% and 120\%} improvement on the overall NDCG and
MAP@R scores, respectively) and provides better root causes compared to its
original version. \response{We also observe that the gaps between \tool\ and
other methods like CauseInfer+CE vary across different settings. For
example, in the Network component, \tool\ outperforms CauseInfer+CE by 66\%
in NDCG. Here, CauseInfer+CE only identifies five causes out of 14, while
\tool\ correctly identifies most of the root causes, which {yield} the
large gap in the Network component. In contrast, in the Database component,
the difference is only 20\%. We find that, {out} of four cases in
{the} Database component, \tool\ provides the best results in three cases
while {being} sub-optimal in the ``Flush Log''.}

\parh{Error Analysis.}~\tool\ has sub-optimal results in a few
settings. We find that \tool\ recommends one \response{KPI} for the anomaly
triggered by memory stress (5th component in \T~\ref{tab:full-comparison}) and
CPU stress (6th component). Therefore, AS is computed using the only KPI. For
memory stress, \tool\ recommends ``Node.Memory\_Distribution\_Free'', which is
voted by all human experts and thus assigned a score of 1. Thus, its Top-1/3/5
AS and NDCG are computed as 1.00. However, experts also annotate some other
useful albeit less important KPIs with lower scores (e.g.,
``MySQL.Query\_Cache\_Mem\_Free\_Mem''). Thus, when a method (e.g.,
CauseInfer) recommends more KPIs besides ``Node.Swap\_Activity\_Swap\_In'',
it gains a higher MAP score. We inspect the cause of this inaccuracy and
find that it is primarily due to inaccurate counterfactual analysis over a
lengthy causal chain (with 11 hops on causal graphs), where regression errors
are accumulated and propagated, resulting in inaccurate estimations.

In another case where an anomaly was triggered by CPU stress (6th component in
\T~\ref{tab:full-comparison}), \tool\ also fails to suggest several highly
relevant KPIs. We find that it is due to the inaccurate causal graph, where the
causal relations between the KPI of interest (i.e., ``Query\_Duration'') and
many CPU-related KPIs are incorrectly missed. Performance of baseline methods
downgrades as well. Given the inherent difficulty of causal structure learning,
we anticipate that \tool\ users will incorporate domain knowledge to revise
certain spurious edges and improve the quality of causal graphs. In particular,
\response{according to domain knowledge and experts' feedback, we notice that
CPU usage is causally related {to} memory and IO activities. Hence, we
manually add four edges from ``Node.CPU\_Usage\_Load\_user'' and
``Node.CPU\_Usage-\_Load\_1m'' pointing to ``Node.Memory\_Distribution\_Free''
and ``Node.Memory\_IO\_Activity''.
After this, \tool's performance is improved from 0.12 to 0.30 on the
MAP@R score and from 0.60 to 0.67 on the NDCG score, respectively.}

\begin{table}[t]
  \centering
  \caption{Statistics of generated causal graphs. \#Node and \#Edge denotes the number of
    non-isolated nodes/edges in the causal graphs. BIC denotes Bayesian Information
    Criterion.}
  \resizebox{0.85\linewidth}{!}{
    \begin{tabular}{l|c|c|c|c}
      \hline
      \textbf{Method} & \textbf{\#Node} & \textbf{\#Edge} & \textbf{BIC} & \textbf{Accuracy} \\
      \hline
      \tool\ w/o CE   & 64              & 63              & -614055.5    & 0.73              \\\hline
      \tool\          & 78              & 89              & -584910.3    & 0.87              \\\hline
      Improvement     & +22\%           & +42\%           & +6\%         & +16\%             \\\hline
    \end{tabular}
  }
  \label{tab:causal-graph-stat}
\end{table}


\parh{Causal Graph Accuracy.}~We report the statistics of causal graphs
generated by \tool\ and its ablated version (\tool\ w/o CE) that excludes the CE
module in \T~\ref{tab:causal-graph-stat}. \textbf{BIC} score is a standard
metric for assessing the ``fitness'' of causal graphs on observational data.
\textbf{Accuracy} is a metric examining {whether} the pairwise correlations are
properly represented in the causal graph in terms of d-separations and
d-connections. We refer readers to~\cite{pgmpy} for the full details of these
metrics. The two metrics are computed using the unseen data collected during the
online phase. For both metrics, a greater value indicates a better fitness with
online data. We interpret the overall results as encouraging: 78 out of 79 KPIs
(after excluding constantly unchanged KPIs) are not isolated in the causal
graph, and 26 additional edges are identified compared to the ablated version.
Furthermore, \tool\ manifests a high degree of agreement with the unseen online
data, with an accuracy of 0.87. The improvements in the BIC score further
{demonstrate} the effectiveness of CE. We supply the full graph in \sm. Overall,
we consider that the \tool's causal discovery can learn an accurate causal
graph, and CE facilitates learning {the} causal graph with much better quality.

\subsection{Evaluation on Scalability}
\label{subsec:tidb}

\parh{Environment.}~We also evaluate \tool\ on a distributed database, TiDB. As
in \S~\ref{sec:implementation}, we host TiDB on a K8s cluster with three
\texttt{TiKV} pods (\texttt{tikv-0,1,2}), one \texttt{TiDB} pod (\texttt{tidb}),
and one \texttt{PD} pod (\texttt{pd}). As noted in \S~\ref{sec:implementation},
254 KPIs are involved in the TiDB scenario, which is much larger than MySQL.
Thus, we view this evaluation as appropriate to benchmark the scalability of
\tool\ on real-world distributed databases. We present two case studies by
imposing two anomalies to TiDB: \texttt{tidb-pod-failure} and
\texttt{network-loss}. \texttt{tidb-pod-failure} injects a failure directly
{into} the \texttt{TiDB} pod and makes it temporally unavailable.
\texttt{tikv-0-network-loss} randomly drops 80\% packets that are sent to a
specific KV pod (i.e., \texttt{tikv-0}). Since the way to trigger performance
anomalies (i.e., the ground truth) is known in our experiments, we employ domain
knowledge to justify whether \tool's outputs are consistent with {the} ground
truth.

\begin{table}[t]
  \centering
  \caption{Top-10 KPIs suggested by \tool\ for answering \texttt{tidb-pod-failure}.}
  \resizebox{1.0\linewidth}{!}{
    \begin{tabular}{c|c|l}
      \hline
      \textbf{Rank}         & \textbf{KPI Family}                 & \textbf{Description}                                      \\
      \hline
      \multirow{3}{*}{1-5}  & \multirow{3}{*}{\makecell{Lock                                                                     \\Resolve OPS}} &
      The number of \texttt{TiDB} operations that resolve                                                                        \\
                            &                                     & locks. When a \texttt{TiDB} pod fails, \texttt{TiDB}-related \\
                            &                                     & operations would decrease.                                   \\\hline
      \multirow{4}{*}{6, 9} & \multirow{4}{*}{\makecell{Statement                                                                \\OPS}} &
      The number of different SQL statements                                                                                     \\
                            &                                     & executed per second by \texttt{TiDB} pod. When a             \\
                            &                                     & \texttt{TiDB} pod fails, it would be unable to               \\
                            &                                     & execute statements.                                          \\\hline
      \multirow{3}{*}{7, 8} & \multirow{3}{*}{\makecell{CPS By                                                                   \\Instance (OK)}} &
      The succeed command statistics on each \texttt{TiDB}                                                                       \\
                            &                                     & instance. When a \texttt{TiDB} pod fails, the number         \\
                            &                                     & of succeed commands would decrease.                          \\\hline
      \multirow{3}{*}{10}   & \multirow{3}{*}{\makecell{KV Cmd                                                                   \\OPS}} &
      The number of executed KV commands emitted                                                                                 \\
                            &                                     & by \texttt{TiDB} pod. When a \texttt{TiDB} pod fails, it     \\
                            &                                     & would be unable to execute statements.                       \\\hline
    \end{tabular}
  }
  \label{tab:tide-case1}
\end{table}

\begin{table}[t]
  \centering
  \caption{Top-10 KPIs suggested by \tool\ for answering \texttt{tikv-0-network-loss}.}
  \resizebox{1.0\linewidth}{!}{
    \begin{tabular}{c|c|l}
      \hline
      \textbf{Rank}            & \textbf{KPI Family}                     & \textbf{Description}                                         \\
      \hline
      \multirow{5}{*}{1, 3, 7} & \multirow{5}{*}{\makecell{\texttt{TiKV}                                                                   \\Write\\Leader}} &
      The number of leaders that are writing on each                                                                                       \\
                               &                                         & \texttt{TiKV} instance. When network packets to a \texttt{TiKV} \\
                               &                                         & pod are constantly lost, writes on this \texttt{TiKV} pod       \\
                               &                                         & would decrease and writes on other pods would                   \\
                               &                                         & increase.                                                       \\\hline
      \multirow{6}{*}{\makecell{2, 4-6,                                                                                                    \\10}} & \multirow{6}{*}{\makecell{\texttt{TiKV}\\Resource}} &
      The usage of resources (e.g., CPU and memory)                                                                                        \\
                               &                                         & on each \texttt{TiKV} instance. When network packets            \\
                               &                                         & to a \texttt{TiKV} pod is constantly lost, it would process     \\
                               &                                         & less requests thus uses less resources, while                   \\
                               &                                         & others are responsible for more requests and                    \\
                               &                                         & use more resources.                                             \\\hline
      \multirow{4}{*}{8, 9}    & \multirow{4}{*}{\makecell{Duration}}    &
      The duration for processing different activities.                                                                                    \\
                               &                                         & When network packets to a \texttt{TiKV} pod is constantly       \\
                               &                                         & lost, the duration of completing different                      \\
                               &                                         & activities would increase.                                      \\\hline
    \end{tabular}
  }
  \label{tab:tide-case2}
\end{table}

\parh{Case 1:}~In \texttt{tidb-pod-failure}, we inject a failure {to} the
\texttt{TiDB} pod. Then, we observe that the duration of the request
launched by \texttt{PD} pod (i.e., ``Handle Requests Duration'') becomes
abnormal and ask \tool\ to diagnose this anomaly. In short, we group the
top-10 KPIs suggested by \tool\ and use TiDB's official documentation to
comprehend each of them. As shown in \T~\ref{tab:tide-case1}, while the
target KPI (PD's request duration) is not directly relevant to \texttt{TiDB}
pod failure, we find that all top-10 KPIs suggested by \tool\ are highly
relevant to the root cause. All of {the} KPIs indicate {the exact}
different activities performed by the failed \texttt{TiDB} pod. Therefore,
we interpret the outputs of \tool\ as reasonable and highly informative:
when users are provided with these KPIs, it should be accurate to assume
that they can easily identify the anomaly's root cause.

\parh{Case 2:}~In \texttt{tikv-0-network-loss}, a notable proportion of
network packets are dropped, reducing the overall performance of the
database. In particular, we observe that the duration of command completion,
``Completed Commands Duration (seconds)'', is abnormal, and we apply \tool\
for root cause analysis. We group and present the results of \tool\ in
\T~\ref{tab:tide-case2}. We find the outputs of \tool\ {to be} valuable.
Note that the affected \texttt{TiKV} pod (\texttt{tikv-0}) can only serve
limited functionality, while other pods (\texttt{tikv-1,2}) would serve
extra {responsibilities}. Therefore, the decrease {in} workload on
\texttt{tikv-0} and the surge {in} workload on \texttt{tikv-1,2} are
reasonable and should provide enough hints for users to investigate the
detailed status of \texttt{tikv-0}.

\subsection{Evaluation on Counterfactual Analysis}
\label{subsec:synthetic}


This section evaluates if \tool\ provides satisfactory counterfactual analysis
results under complex causal graphs. Collecting ground-truth outcomes of
quantitative counterfactual changes in real-world DBMSs is challenging, if not
impossible (hence the human evaluation in \S~\ref{subsec:mysql}). Therefore, we
use synthetic data, a common setup for evaluating causal inference algorithms,
to compare \tool\ and other baselines.  

\begin{table}[t]
  \centering
  \caption{DGP models for generating synthetic data. LS stands for local structure.}
  \resizebox{1.0\linewidth}{!}{
    \begin{tabular}{c|l}
      \hline
      \textbf{LS (see \F~\ref{fig:local-structure})} & \textbf{Formulation}                                                            \\
      \hline
      \multirow{3}{*}{\makecell{(a)                                                                                                    \\no confounder}}  & $X_1\sim \text{Uniform}(-1,1)$ \\
                                                     & $X_2\sim \text{Uniform}(-1,1)$                                                  \\
                                                     & $Y=\theta_1(X_1) + \theta_2(X_2) + \epsilon, \epsilon\sim \text{Uniform}(-1,1)$ \\\hline
      \multirow{3}{*}{\makecell{(b)                                                                                                    \\observable confounder}}  & $X_1\sim \text{Uniform}(-1,1)$ \\
                                                     & $X_2=\theta_1(X_1)+\eta,\eta\sim \text{Uniform}(-1,1)$                          \\
                                                     & $Y=\theta_2(X_1) + \theta_3(X_2) + \epsilon, \epsilon\sim \text{Uniform}(-1,1)$ \\\hline
      \multirow{4}{*}{\makecell{(c)                                                                                                    \\latent confounder}}  & $IV\sim \text{Uniform}(-1,1)$ \\
                                                     & $X_1\sim \text{Uniform}(-1,1)$ (\underline{unobserved})                         \\
                                                     & $X_2=\theta_1(X_1)+\theta_2(IV)+\eta,\eta\sim \text{Uniform}(-1,1)$             \\
                                                     & $Y=\theta_3(X_1) + \theta_4(X_2) + \epsilon, \epsilon\sim \text{Uniform}(-1,1)$ \\\hline
    \end{tabular}
  }
  \label{tab:local-structure}
\end{table}


\parh{DGP Model.}~Following common practice in causality
analysis~\cite{econml,zheng2018dags}, we formulate a set of DGP (Data Generating
Process) models~\cite{econml} for evaluation in \T~\ref{tab:local-structure}.
These models correspond to the local structures (with confounders) in
\S~\ref{subsec:offline}. We use a linear form for $\theta$ and define all
exogenous variables to follow a uniform distribution. Then, we generate $100$
datasets for each DGP model using different random states, resulting in 300
($3\times 100$) DGP models. For each model, we train with 5,000 data samples and
create 1,000 counterfactual queries. Each query answers the treatment effects of
changing $X_2=x_2$ to $X_2=x_2'$ when $X_1=x_1$, where $x_1,x_2,x_2'$ are
generated randomly and the ground-truth treatment effects are generated by the
specification of $Y$ in DGP models.


\parh{Baseline \& Metric.}~We use Multi Layer Perception (MLP), Random Forest
(RF), Decision Tree with Pruning (DTP), and Support Vector Machine (SVM) as
baseline methods, treating counterfactual analysis as a regression problem. The
MLP model has three hidden layers and uses the Adam
optimizer~\cite{kingma2014adam} with default parameters. We use default
parameters for all models and report MSE (mean squared error; lower is better)
and $R^2$ (coefficient of determination; higher is better) for each method on
datasets generated by each DGP model.

\begin{table}[t]
  \centering
  \caption{MSE and $R^2$ on synthetic data. LS stands for local structure. Best
    metrics are \colorbox{green}{highlighted}.}
  \resizebox{1.0\linewidth}{!}{
    \begin{tabular}{c|c|c|c|c|c|c|c}
      \hline
      \textbf{LS}          & \multicolumn{2}{|c|}{\textbf{Metric}} & \textbf{\tool} & \textbf{MLP}      & \textbf{RF} & \textbf{DTP} & \textbf{SVM}                     \\\hline
      \multirow{4}{*}{(a)} & \multirow{2}{*}{MSE}                  & mean           & \bestcell{0.0003} & 0.0020      & 0.0109       & 0.0634       & \bestcell{0.0003} \\\cline{3-8}
                           &                                       & std            & \bestcell 0.0003  & 0.0011      & 0.0020       & 0.0228       & 0.0005            \\\cline{2-8}
                           & \multirow{2}{*}{$R^2$}                & mean           & \bestcell 0.9988  & 0.9914      & 0.9522       & 0.7422       & 0.9983            \\\cline{3-8}
                           &                                       & std            & \bestcell 0.0018  & 0.0063      & 0.0297       & 0.1456       & 0.0029            \\\hline
      \multirow{4}{*}{(b)} & \multirow{2}{*}{MSE}                  & mean           & \bestcell 0.0002  & 0.0030      & 0.0176       & 0.1206       & 0.0003            \\\cline{3-8}
                           &                                       & std            & \bestcell 0.0003  & 0.0022      & 0.0072       & 0.0549       & 0.0005            \\\cline{2-8}
                           & \multirow{2}{*}{$R^2$}                & mean           & \bestcell 0.9988  & 0.9875      & 0.9282       & 0.4949       & 0.9983            \\\cline{3-8}
                           &                                       & std            & \bestcell 0.0019  & 0.0093      & 0.0463       & 0.3501       & 0.0028            \\\hline
      \multirow{4}{*}{(c)} & \multirow{2}{*}{MSE}                  & mean           & \bestcell 0.0174  & 0.3758      & 0.1029       & 0.0815       & 0.0389            \\\cline{3-8}
                           &                                       & std            & \bestcell 0.0302  & 0.2098      & 0.0960       & 0.0337       & 0.0289            \\\cline{2-8}
                           & \multirow{2}{*}{$R^2$}                & mean           & \bestcell 0.9382  & -0.8258     & 0.1540       & 0.5733       & 0.7797            \\\cline{3-8}
                           &                                       & std            & \bestcell 0.0936  & 1.9738      & 1.2979       & 0.3894       & 0.2619            \\\hline
    \end{tabular}
  }
  \label{tab:synthetic-result}
\end{table}


\T~\ref{tab:synthetic-result} shows that \tool\ excels in almost all settings.
While SVM has comparable MSE performance in LS (a), this is reasonable since
\tool\ and SVM are asymptotically equivalent without confounders. However, SVM's
performance declines in LS (b) and (c). Specifically, in LS (c), the MLP model
overfits the data, resulting in unsatisfactory and unstable performance. This
downgrade is reasonable, as the MLP model cannot distinguish the effects of
$X_1,X_2$ on $Y$ from the data generated by DGP models under LS (c). Other
models exhibit similar downgrades. In contrast, \tool\ accurately estimates the
effects of $X_2$ on $Y$ across all settings, even with latent confounders, as in
LS (c).

\section{Discussion and Threats to Validity}
\label{sec:discussion}

\parh{Capturing Temporal Characteristics of KPIs.}~Like most related
works~\cite{chen2014causeinfer,lin2018microscope,jeyakumar2019explainit}, this
research assumes KPI observations are independent and identically distributed
(iid), making causality analysis feasible. However, KPIs may exhibit temporal
characteristics in reality, potentially leading to missed edges in causal
structure learning and inaccurate counterfactual analysis estimations. As causal
discovery and inference on time series remain largely understudied, we defer
considering the temporal properties of KPIs in counterfactual analysis to future
research.

\parh{Enhancing CE.}~We recognize that the current usage of \cm\ is suboptimal,
particularly due to redundancy among CE experiments. Minimizing the number of
experiments while maximizing causality analysis observability presents a
challenging problem for future exploration. We expect that certain
``feedback-driven'' mutations (analogous to feedback-driven fuzz
testing~\cite{afl}) can be applied when mutating chaos variables, reducing
invocations of \cm.

\parh{Threats to Validity.}~One threat is that the evaluation of \tool\ is
limited to two real-world databases and our conclusion may not generalize to
other databases. Another relevant threat is on the potentially limited set of
performance anomalies, as we mainly focus on database response time. We mitigate
these threats by designing \tool\ whose technical pipeline is independent of
specific databases or response time. We envision that \tool\ is applicable to
other scenarios like cloud containerization systems, and \tool\ can diagnose
other performance anomalies like database throughput.

\section{Related Work}
\label{sec:relatedwork}

\parh{Theoretical Causality Analysis.}~Causality analysis is
the foundation of scientific discovery, with applications in the fields of
economics and clinical trials~\cite{pearl2009causality}. Identifying
qualitative (i.e., SEM structure learning) and quantitative (i.e., SEM
parameter learning) causal relationships is challenging. Typically, causal
discovery algorithms are applied to learn a causal graph that encodes
qualitative causal relationships among a set of variables. Recent years
have witnessed a surge of interest in causal
discovery~\cite{scanagatta2015learning, zheng2018dags, ding2020reliable,
ma2021mt, lorch2021dibs, ma2022ml4s, ma2022noleaks}. Without clear
qualitative causal relationships, causal inference (i.e., identifying
quantitative causal relationships) is infeasible or inaccurate. The methods
for causal inference are diverse, including actual
causality~\cite{halpern2020causes} and potential outcomes
framework~\cite{rubin2005causal, rubin1974estimating} proposed by Rubin.
\tool\ uses Rubin's framework because it is more suitable when KPIs are
modeled as random variables and the ATE (average treatment effect)
naturally allows for propagating influence over causal paths.

\parh{Applications of Causality Analysis.}~Recently, there
have been a number of researches that applied causality analysis to address
problems in software engineering~\cite{fariha2020causality,
dubslaff2022causality, johnson2020causal, krishna2020cadet,
hsiao2017asyncclock}. These tools usually focus on analyzing how program
inputs or configurations impact the software behavior (e.g., execution time
or crash). In addition, causality analysis is used extensively in the field
of machine learning due to its inherent
interpretability~\cite{sun2022causality, zhang2022adaptive, ma2022xinsight,
ji2023cc, MonjeziTTT23, ji2023causality}. We also believe that causality
analysis can have a broader application in software engineering because it
can systematically handle interactions among many variables; we will
explore the usage of causality analysis in other fields like testing and
static analysis in the future.

\section{Conclusion}

We have presented \tool\ for database performance anomaly diagnosis. \tool\
novelly employs \cm\ for augmenting causality-based performance debugging, and
it features both qualitative root cause identification and quantitative
counterfactual analysis and addresses several challenges in the process.
Evaluations on synthetic and real scenarios show that \tool\ offers accurate
diagnosis and outperforms existing works across nearly all settings. 
\ifCLASSOPTIONcaptionsoff
  \newpage
\fi




\bibliographystyle{IEEEtran}
\bibliography{bib/ref}

%
%
%
%
\end{document}